\documentclass[twocolumn,11pt]{article}
\usepackage{times}
\usepackage{graphicx}
\newtheorem{definition}{Definition}

\newtheorem{lemma}[definition]{Lemma}
\newtheorem{theorem}[definition]{Theorem}
\newtheorem{proposition}[definition]{Proposition}
\newtheorem{corollary}[definition]{Corollary}
\newcommand{\R}{{\rm I}\!{\rm R}} 
\newcommand{\N}{{\rm I}\!{\rm N}} 

\begin{document}
\global\def\refname{{\normalsize \it References:}}
\baselineskip 12.5pt
%
%
%
\title{\LARGE \bf Some Proofs on Statistical Magnitudes for Continuous Phenomena}

\date{}

\author{\hspace*{-10pt}
\begin{minipage}[t]{2.0in} \footnotesize \baselineskip 10.5pt
\centerline{\normalsize RAQUEL G. CATAL\'AN}
\centerline{Universidad P\'ublica de Navarra}
\centerline{Department of Mathematics}
\centerline{E-31006 Pamplona}
\centerline{SPAIN}
\centerline{raquel.garcia@unavarra.es}
\end{minipage}
\begin{minipage}[t]{2.0in} \footnotesize \baselineskip 10.5pt
\centerline{\normalsize JOS\'E GARAY}
\centerline{Universidad de Zaragoza}
\centerline{Department of Mathematics}
\centerline{E-50009 Zaragoza}
\centerline{SPAIN}
\centerline{jgaray@unizar.es}
\end{minipage}
\begin{minipage}[t]{2.0in} \footnotesize \baselineskip 10.5pt
\centerline{\normalsize RICARDO L\'OPEZ-RUIZ}
\centerline{Universidad de Zaragoza}
\centerline{Department of Computer Science and BIFI}
\centerline{E-50009 Zaragoza}
\centerline{SPAIN}
\centerline{rilopez@unizar.es}
\end{minipage} 
%
%
\\ \\ \hspace*{-10pt}
\begin{minipage}[b]{6.9in} \normalsize
\baselineskip 12.5pt {\it Abstract:}
{\small In this work, the proofs concerning the continuity of 
the disequilibrium, Shannon information and statistical complexity
in the space of distributions are presented. Also, some results on the
existence of Shannon information for continuous systems are given.} 
\\ [4mm] {\it Key--Words:}
Near-continuity, disequilibrium, Shannon information, statistical complexity
\end{minipage}
\vspace{-10pt}}

\maketitle

\thispagestyle{empty} \pagestyle{empty}
%
%

\section{Introduction}
\label{S1} \vspace{-4pt}

The concepts of Shannon information (or entropy), $S$, and disequilibrium, $D$, 
for a continuous system are given by
 \begin{eqnarray}
 S & = & - k\, \int_{-\infty}^{\infty} p(x)\, \log\, p(x)\, dx\,, \\
 D & = & \int_{-\infty}^{\infty} p^2(x)\, dx\,, 
 \end{eqnarray}
where $k$ is a positive constant (that is chosen to be equal to 1),
$x$ represents the continuum of the system states,
and $p(x)$ stands for the normalized density function of all these states.

The statistical complexity $C$, the so-called $LMC$ complexity \cite{lopez1995}, 
is defined as 
\begin{equation}
C = H\cdot D\;,
\end{equation}
where $H$ gives account of the broadness of the distribution and
and $D$ gives an idea of how much spiky is the distribution respect to equilibrium distribution,
that in this case is the equiprobability. 
For our purpose, we take a version used in Ref. \cite{garay2002}
as quantifier of $H$. This is the simple exponential Shannon entropy \cite{dembo1991},
that takes the form, 
\begin{equation}
H = e^{S}\;,
\end{equation}
that implies the positivity of $H$ and $C$. These information theoretic indicators
has been successfully applied in different contexts in order to unveil a complex behavior or structure,
for instance, in gases out of equilibrium \cite{calbet2001}, in coupled map lattices \cite{sanchez2005}
or in quantum systems \cite{sanudo2008}, to cite some of them.

In this communication, it is our objective to prove rigorously that all these functionals, 
$H$, $D$ and $C$, are near-continuous, i.e. similar distributions take a similar value of 
information, disequilibrium and statistical complexity. Also, we present some exact results
on the values that Shannon information can take for continuous systems.

 \section{Continuity of $H$, $D$ and $C$}
 \label{S2} \vspace{-4pt}

 First, we remind some definitions that were used in Ref. \cite{garay2002}.

 \begin{definition}.
Let $I$ be an interval in $\R$ and $\delta >0$. We will say that two density functions, 
namely $f$ and $g$, are $\delta$-neighboring functions on $I$ if both are supported 
on $I$ and the Lebesgue measure of the set of points of $I$ such that $|f(x)-g(x)|\ge \delta$ 
is zero, i.e. if the essential supremun (ess sup) of $|f-g|$ verifies
 $$
 {\rm ess}\sup_{x\in I} |f(x) - g(x)|  < \delta.
 $$
 \end{definition}
 Note that this definition can be applied to non bounded density functions.

\begin{definition}.
 We will say that  a real map $T$ defined on density functions
 is near-continuous on $I$ if for any $ \varepsilon >0$ there exists
 $\delta (\varepsilon, I) >0$ such that if
 $f$ and $g$ are two $\delta$-neighboring functions on $I$  then
 $|T(f) - T(g)| <\varepsilon$.
\end{definition}
 
\begin{theorem}.
\label{th1}
 Disequilibrium $D$ is near-continuous on $\R$.
\end{theorem}

{\bf Proof}.
Take $\varepsilon >0$, $\delta < {\varepsilon \over 2}$. Let $f$, $g$ be two 
$\delta$-neighboring functions on $\R$. Then,
 \begin{eqnarray}
 D(f) - D(g) & = & \int_{-\infty}^{\infty} (f^2(x) - g^2(x) )\, dx  \\
 & = & \int_{-\infty}^{\infty} (f(x) + g(x))* \nonumber \\
  & &  \;\;\;\;\;\;\; (f(x) - g(x))\, dx\,. \nonumber
 \end{eqnarray}
As $f$, $g$ are $\delta$-neighboring functions on 
$\R$ we have that $|f(x) - g(x)| \le \delta $ a.e. and so,
 \begin{eqnarray*}
 |D(f) - D(g)| & \le & \int_{-\infty}^{\infty} (f(x) + g(x))* \\
  & & \;\;\;\;\;\;\; |f(x) - g(x)|\,dx \nonumber \\
  & \le &  \delta\, \left[ \int_{-\infty}^{\infty} f(x)\,dx + \right. \nonumber \\
  & & \;\;\;\;\;\;\; \left.\int_{-\infty}^{\infty} g(x)\,dx \right] \nonumber \\
  & = &  2\delta < \epsilon \,. \hspace{0.5cm}\clubsuit
 \end{eqnarray*}

Note that if we consider only bounded density functions, then $D$ is a
continuous map between normed spaces in the topological sense.
  
In order to prove the near-continuity of the Shannon information, $S$,
and for extension of its exponential, in this case, $H$, 
the following technical Lemma is needed.

\begin{lemma}.
\label{lm1}
For all $\alpha>1$ there exists
 $\delta_{\alpha} < 1$  such that for any pair of points $x_1$, $x_2$ with
 $0<x_1 ,x_2 < \alpha$ and $|x_1 - x_2|<\delta <\delta_{\alpha}$ we have that
 $$
 |x_1 {\rm log}x_1 - x_2 {\rm log}x_2| < |\delta {\rm log}\delta|\,.
 $$
\end{lemma} 
 
 {\bf Proof}.
 Given $\delta \in (0,1)$ consider the function
 $$
 s_{\delta} (x) = (x+\delta ) {\rm log}(x+\delta) - x {\rm log}x,
 $$
which can be extended to $x=0$ with
continuity and is derivable in $\R^{+}$ with $s_{\delta}^{\,'}(x) = {\rm log}{ x +\delta \over x}$.
As $\delta >0$, $x+\delta >x$ and so $s_{\delta}^{\,'}>0$. This means that
$s_{\delta}$ is rising on $\R^+$.
Observe that as $\delta <1$, ${\rm log} \delta <0$ and so $s_{\delta}(0) <0$. 
On the other hand, as $s_{\delta}$ is rising and $\alpha>1$, we have that 
$0 < (1+\delta) {\rm log} (1+\delta ) = s_{\delta} (1) < s_{\delta}(\alpha)$. 
This means that there is only one unique $x_0$ so that $s_{\delta} (x) <0$ for any $0\le x < x_0$ 
and $s_{\delta} (x)>0$ for the values of $x_0 < x \le \alpha$.

The result of the Lemma, expressed in terms of the function $s_{\delta}$,
is equivalent to finding $\delta_{\alpha}$ such that for any $0<\delta<\delta_{\alpha}$ 
and $0<x<\alpha$,
 $$
 |s_{\delta}(x)| <  |s_{\delta}(0)|.
 $$

Clearly, this inequality is verified for any $x < x_0$. So, we can suppose $x>x_0$. 
In this case the inequality can be written as
$$
s_{\delta} (x) < -s_{\delta} (0).
$$

On the other hand, as $s_{\delta}$ is rising, it can be seen that the verification of 
the former inequality is equivalent to the verification of the next one:
$$
s_{\delta} (\alpha ) < - s_{\delta}(0).
$$

Let us consider the function
$F_{\alpha} (\delta ) = s_{\delta} (\alpha ) + s_{\delta}(0)$.
The problem has been reduced to finding a $\delta_{\alpha}$ such
that for all $0<\delta<\delta_{\alpha}$,
$F_{\alpha} (\delta ) <0$.
Since  $F'_{\alpha}(\delta) = 2 + {\rm log}(\delta(\alpha +\delta))$, 
we will take $\tilde\delta_{\alpha}$ as the only positive number such that
 $F'_{\alpha}(\delta) =0$, then
 $$
 \tilde\delta_{\alpha} =
 {1\over {2 e}} \left(-\alpha e + \sqrt{(\alpha e)^2 +4}\right).
 $$
As $F_{\alpha}$ is decreasing on $[0, \tilde\delta_{\alpha}]$, 
rising on $[\tilde\delta_{\alpha}, \infty)$ and $F_{\alpha}<0$, we
can conclude that there exists an unique $\delta_{\alpha}$ bigger than 
$\tilde\delta_{\alpha}$ such that for all $0<\delta<\delta_{\alpha}$, 
$F_{\alpha}(\delta) <0$. $\hspace{0.5cm}\clubsuit$

\begin{theorem}.
\label{th2}
$H$ is near-continuous on every compact $K\subset \R$.
\end{theorem}
 
{\bf Proof}.
First, we will prove that $S$ is near-continuous.
Then, by continuity of the exponential operation, $H$ will also be near-continuous.

Let $K$ be a compact set in $\R$. Take $\varepsilon<1$, $\delta>0$ with
$\delta < {\rm min}\{\delta_2, {\varepsilon \over 8}\}$, where $\delta_2$ is
the $\delta_\alpha$ given in  Lemma \ref{lm1} for $\alpha=2$, and
$|\delta {\rm log}\delta |< {\varepsilon \over {2|K|}}$, where $|K|$ stands
for the Lebesgue measure of $K$.

Let $f$ and $g$ be two $\delta$-neighboring functions on $K$ and let us
consider the sets
\begin{eqnarray*}
 A & = & \{ x\in K;\, f(x)>g(x)\ge 1\},\\
 B & = & \{ x\in K;\, g(x)>f(x)\ge 1\},\\
 C & = & \{ x\in K;\, f(x)=g(x)\},\\
 D & = & \{ x\in K;\, f(x)\le 2,\, g(x)\le 2\}\,.
\end{eqnarray*}

 As $\delta<1$, $A \cup B \cup C \cup D = \R$. Defining 
 $G = f\log f - g\log g$, we have
 \begin{eqnarray}
 \label{Hineq}
 |S(f) - S(g)|  & \le & \int_{-\infty}^{\infty} |G(x)|\, dx  \\
 & \le &  \int_A |G(x)|\, dx + \int_B |G(x)|\, dx + \nonumber \\
 &     &  \int_C |G(x)|\, dx + \int_D |G(x)|\, dx  \, . \nonumber
 \end{eqnarray}

We will bound separately each one of the above integrals. Note that the
third one is null and the first two ones behave in a similar manner.
 
Applying the Mean Value Theorem to the function
$s(t) = t\,\log(t)$ we have that for any $x\in A$ there exists $\tilde x$ 
such that $g(x)<\tilde x<f(x)$ and
 \begin{eqnarray*}
 G(x) & = & [f(x) - g(x)] s' (\tilde x) \\
  & = & [f(x) - g(x)][1 + {\rm log}\tilde x]\,.
 \end{eqnarray*}

As $x\in A$, $1\le g(x)$. Then, $1\le \tilde x < f(x)$ and so 
$\log \tilde x < \tilde x < f(x)$. From here we conclude that $1 + \log\tilde x <f(x) + g(x)$ 
and then the first integral can be bounded as
 \begin{eqnarray*}
 \int_A |G(x)|\, dx & \le & \int_{A} |f(x) - g(x)| [1+ \log \tilde x] \, dx \\
 & \le & \delta \, \int_{\R} (f(x) +g(x)) \, dx = 2\delta \, < {\varepsilon \over 4}\,. 
 \end{eqnarray*}

Analogously,
 $$
 \int_B |G(x)|\, dx \, < {\varepsilon \over 4}\,.
 $$

For the last integral we make use of the Lemma \ref{lm1}. 
As $\delta < \delta_2$, we have that
 $$
 |G(x)| = |f\log f - g\log g|
      < |\delta {\rm log}\delta |.
 $$
And as $\delta$ has been chosen so that $|\delta \log\delta| < {\varepsilon \over {2 |K|}}$
 $$
 \int_D |G(x)| \, dx \le {\varepsilon \over 2}\,.
 $$

By putting together in (\ref{Hineq}) the inequalities obtained for each integral,
then finally it is proved that $S$ is near-continuous. Hence, by extension,
$H$ is also near-continuous and the Theorem \ref{th2} is proved. $\hspace{0.5cm}\clubsuit$
 
\begin{corollary}.
 The complexity $C$ is near-continuous on compacts.
\end{corollary}

{\bf Proof}.
 It is immediate from the definition of $C$ and  Theorems \ref{th1} and
\ref{th2}. $\hspace{0.5cm}\clubsuit$ 

As it was observed in Ref. \cite{garay2002}, let us remark the necessity to have
compactness in the sets $K$ in order to prove the near-continuity of $H$.
The same can be said for $C$.

\section{On the Existence of Entropy}
\label{S3} \vspace{-4pt}

Here, we proceed to show some results on the existence of the Shannon entropy $S$
for continuous systems.

\begin{proposition}.
Given a density function $p(x)$ bounded and compactly supported, its
entropy $S(p)$ is finite.
\end{proposition}

{\bf Proof}.
Let $C$ be the upper bound of $p(x)$. 
As $0< p\le C$ on the support $K$ of $p$, and the function $x \log x$ has range 
in $[-e^{-1},\infty)$ for $x>0$, then
$$
-{1\over e}\,\mu (K)\le -S(p)\le max\{0,C\log C\} \,\mu (K)\,.
$$
So the integral of the entropy is convergent. $\hspace{0.5cm}\clubsuit$

Let us observe that if $p(x)$ is not bounded or has no bounded support, this proposition may not 
be truth. In this situation, it will be possible to have one of the following cases:
\begin{itemize}
\item{(1)} $S(p)$ does exist and is finite,
\item{(2)} $S(p)$ does exist and is $+\infty$,
\item{(3)} $S(p)$ does exist and is $-\infty$,
\item{(4)} $S(p)$ does not exist.
\end{itemize}

All these cases are possible, as it can be seen with the following example. 
Take $p(x)={1\over x}$ supported on different sets in $\R^+$. 
Without loss of generality it will be assumed that the support of the 
density function $p(x)$ is a set $B$  consisting on
a countable union of intervals in $\R^+$.

The following notation: 
$B=B^+\cup B^- = \left(\bigcup _{n\in\N} I_n \right)\cup\left( \bigcup _{n\in\N } J_n\right)$ 
is used, where
$$
I_n=(\gamma_n,\gamma'_n) \hspace{0.5cm}\hbox{verifies} 
$$
$$
\hskip 20 pt 1\le \gamma_0\le \gamma'_0\le\dots\le\gamma_n\le\gamma'_n\le\dots
$$
$$
\hbox{and} \hspace{0.5cm} J_n=(\eta_n,\eta'_n) \hspace{0.5cm}\hbox{verifies} 
$$ 
$$
\hskip 20 pt 1\ge \eta'_0\ge \eta_0\ge\dots\ge\eta'_n\ge\eta_n\ge\dots > 0\,.
$$

Now we consider the projection of such intervals onto $\R$ through the logarithmic function. 
Let us call  $T_n=(\tau_n,\tau'_n)=\log I_n$ and $S_n=(\sigma_n,\sigma'_n)=\log J_n$ and 
so $\dots\le \sigma_n\le\sigma'_n\le\dots\le\sigma_0\le \sigma'_0\le 0\le
\tau_0\le\tau'_0\le \dots\le\tau_n\le\tau'_n\le\dots$. 
We will call $A= \left(\bigcup_{n\in\N} T_n \right)\cup\left(\bigcup_{n\in\N} S_n\right)=A^+ \cup A^-$,
and introduce the notation $\delta_n =\tau'_n-\tau_n$ and $\rho_n =\sigma'_n -\sigma_n$.

It is easy to see that  $p(x)={1\over x}\chi_B(x)$ is of density $1$ if and only if $\sum_{n=0}^\infty
\left(\delta_n +\rho_n\right)=1$.

Let $\mu$ be the Lebesgue measure. Let us define
$$
f(x)=\cases{ \;\;\;\mu \{y>x; y\in A\}   & $x>0$, \cr 
                  -\mu \{y<x; y\in A\} & $x<0$,\cr }
$$
and observe that 
$$
\lim_{x\to 0^+} f(x)-\lim_{x\to 0^-} f(x)=\mu (A^+)+\mu (A^-)=1\,.
$$

\begin{theorem}. 
Let the former $p(x)={1\over x}\chi_B(x)$ be a density function. Then,
\begin{enumerate}
\item If $f$ is integrable Riemann on $\R$ ($f\in {\cal R}(\R)$) then $S(p)\in \R $.
Besides, the finite value of $S(p)$ is given by
$$
S(p)=\int_{-\infty}^{\infty} f(x) dx = \int_{\R^-} f(x) dx + \int_{\R^+} f(x) dx\,.
$$
\item If the value of the integral of $f$ in $\R^-$ is finite and in $\R^+$ is infinite,
i.e. $f\in {\cal R}(\R^-)\setminus {\cal R}(\R^+)$, then $S(p)=+\infty$\,.
\item If the value of the integral of $f$ in $\R^-$ is infinite and in $\R^+$ is finite,
i.e. $f\in {\cal R}(\R^+)\setminus {\cal R}(\R^-)$, then $S(p)=-\infty$\,.
\item If $f\not\in {\cal R}(\R^-)\cup {\cal R}(\R^+)$ then $S(p)$ does not exist\,.
\end{enumerate}
\end{theorem}

{\bf Proof}.
First, we see that
\begin{equation}
\label{eq11}
-\int_{1}^\infty p(x) \log p(x) dx = \int_0^\infty f(x) dx\,. 
\end{equation}
In fact,
\begin{eqnarray*}
 \int_1^{\infty} p(x)\log p(x) dx & = & -\int_{B^+} {\log (x)\over x} \,dx \\
 & = & -\sum_{n=0}^{\infty} \int_{I_n} {\log (x)\over x} \,dx \\ 
 & = & -{1\over 2}\sum_{n=0}^{\infty} \log^2 (x) |_{I_n} \\
 & = & -{1\over 2}\sum_{n=0}^{\infty} ({\tau'}_n^2 - \tau_n^2) \\
 & = & -{1\over 2}\sum_{n=0}^{\infty} \delta_n (\tau'_n+\tau_n)\,. 
\end{eqnarray*}

By other side, analyzing the function $f$ on $\R^+$, we have that 
$\forall  x\in [0, \tau_0]$, $f(x)=\mu(A^+)$, and that $\forall x\in
[\tau'_0,\tau_1]$, $f(x)=\mu (A^+)-\delta_0$. Also, 
$\forall x\in [\tau_0,\tau'_0]$, $f(x)=\mu (A^+)-(x-\tau_0)$, and so on. 
It is also known that $\sum_{n\in \N} \delta_n =\mu (A^+)$. Hence,
\begin{eqnarray*}
\int_0^\infty f(x) dx & = & \sum_{n=0}^\infty (\delta_n \tau_n +{1\over 2}\delta_n^2) \\
& = & \sum_{n=0}^\infty \delta_n (\tau_n + {1\over 2}\delta_n)\\ 
& = & \sum_{n=0}^\infty (\delta_n (\tau_n +{1\over 2}(\tau'_n-\tau_n))\\ 
& = & {1\over 2}\sum_{n=0}^\infty \delta_n (\tau'_n +\tau_n) \,.
\end{eqnarray*}

This proves (\ref{eq11}).

Similarly, it can be proved that
\begin{equation}
\label{eq22}
-\int_{0}^1 p(x) \log p(x) dx = \int_{-\infty}^0 f(x) dx. 
\end{equation}

The result of the theorem is obtained immediately from (\ref{eq11}) and (\ref{eq22}).
$\hspace{0.5cm}\clubsuit$

\section{Conclusion}
\label{S4} \vspace{-4pt}

Continuity is a desirable property for any good behaved magnitude,
although it does not imply the triviality of its mathematical proof.
Here, we have shown that entropy, disequilibrium and statistical complexity
are continuous magnitudes when applied to continuous phenomena.
In this case, it has been put in evidence that the property of compactness 
for the support of the density distribution is crucial.
Also, some exact results on the possible values that entropy can take have been presented.

\vspace{10pt} \noindent
{\bf Acknowledgements:} R. L-R. acknowledges some financial support 
by the spanish Grant with Ref. FIS2009-13364-C02-C01.

\end{document}